\documentclass[fleqn,preprint,5p]{elsarticle}

\usepackage{amsmath,graphicx,amssymb}
\usepackage{epsf,psfig}

\journal{New Astronomy}

\begin{document}

\begin{frontmatter}

\title{Using the $S(c)$ spectrum to distinguish between order and chaos in a 3D \\
galactic potential}

\author{Nicolaos D. Caranicolas}
\author{Euaggelos E. Zotos\corref{}}

\address{Department of Physics, \\
Section of Astrophysics, Astronomy and Mechanics, \\
Aristotle University of Thessaloniki \\
GR-541 24, Thessaloniki, Greece}

\cortext[]{Corresponding author: \\
\textit{E-mail address}: evzotos@astro.auth.gr (Euaggelos E. Zotos)}

\begin{abstract}
The regular and chaotic character of orbits is investigated in a 3D potential describing motion in the central parts of a barred galaxy. This potential is an extension in the 3D space of a 2D potential based on a family of figure-eight orbits, which was produced using the theory of the inverse problem. Starting from the results obtained from the 2D system, we proceed to locate the regions in the phase space of the 3D potential, producing regular or chaotic orbits. In order to obtain this we use a new dynamical parameter, the $S(c)$ spectrum, which proves to be a useful and fast indicator in order to distinguish regular motion from chaos in 3D potentials. Comparison with other methods for detecting chaos is also discussed.
\end{abstract}


\end{frontmatter}

\section{Introduction}

In the present paper we shall study the regular or chaotic character of orbits in the 3D galactic potential
\begin{flalign}
&V(x,y,z) = \frac{1}{2}\left(x^2 + 4y^2 + z^2\right)& \nonumber \\
&- \epsilon\left[\left(1 - x^2\right)^2 + \left(6x^2 - 5\right)y^2 + \alpha\left(x^2 + y^2\right)z^2\right],&
\end{flalign}
where $\alpha$ is a parameter, while $\epsilon$ is the perturbation strength. In all cases we use the value $\alpha = -0.5$. Potential (1) is an extension to the 3D space of the potential obtained using a family of figure-eight orbits and the theory of the inverse problem of dynamics (IPD). The reader can find more details on this subject in Caranicolas (1998). Furthermore, potential (1) can be considered to describe the motion in the central parts of a barred galaxy.

Our aim is to find the regular or chaotic character of orbits in the 3D potential (1). Particular interest will be given to the evolution of the figure-eight orbits, which are the building blocks of the 2D potential. In order to do this, we shall use the new dynamical parameter introduced recently, that is the $S(c)$ spectrum. This dynamical parameter was introduced in a recent paper (see Caranicolas \& Papadopoulos, 2007) and was successfully used to distinguish between regular and chaotic orbits in 2D dynamical systems. In this article, we shall prove that the $S(c)$ spectrum can be also used as a useful tool to distinguish between regular and chaotic motion in 3D potentials.

In a clockwise rotating frame with angular velocity $\Omega_b$, the equations of motion read
\begin{flalign}
&\ddot{x} = - \frac{\partial \ V_{eff}}{\partial x} - 2\Omega_b\dot{y},& \nonumber \\
&\ddot{y} = - \frac{\partial \ V_{eff}}{\partial y} + 2\Omega_b\dot{x},& \nonumber \\
&\ddot{z} = - \frac{\partial \ V_{eff}}{\partial z},&
\end{flalign}
where the dot indicates derivative with respect to the time, while
\begin{flalign}
&V_{eff}\left(x,y,z\right) = V\left(x,y,z\right) - \frac{1}{2}\Omega_b^2\left(x^2 + y^2\right),&
\end{flalign}

The equations of motion (2) admit the integral of motion
\begin{flalign}
&H_J = \frac{1}{2}\left(p_x^2 + p_y^2 + p_z^2\right) + V_{eff}\left(x,y,z\right) = E_J,&
\end{flalign}
which is the Jacobi integral. Here $p_x$, $p_y$ and $p_z$ are the momenta per unit mass conjugate to $x$, $y$ and $z$ respectively, while $E_J$ is the numerical value of the Jacobi integral.

The results of this research are based on the numerical integration of the equations of motion (2), which was done using a Bulirsh-St\"{o}er routine in double precision. The accuracy of the calculations was checked by the constancy of the Jacobi integral, which was conserved up to the twelfth significant figure.

The article is organized as follows: In Section 2 the different families of orbits are presented. In the same section we study the character of orbits in the 2D potential. In Section 3 we investigate the behavior of the 3D orbits using the $S(c)$ spectrum. Special interest is given to the study of the 3D figure-eight orbits as these orbits form the backbone of the whole set of orbits in the 2D galactic potential. Section 4 is devoted to a discussion and the conclusions of this research.

\section{Orbits and spectra in the 2D potential}

In this Section we shall study the character of orbits in the 2D potential. In this case we set $z=0$ and $p_z=0$ in (4) and the corresponding 2D Jacobi integral becomes
\begin{flalign}
&H_{J2} = \frac{1}{2}\left(p_x^2 + p_y^2\right) + V_{eff}\left(x,y\right) = E_{J2},&
\end{flalign}
where $E_{J2}$ is the numerical value of the Jacobi integral. Now the phase space of the system is four dimensional and we can use the classical method of the $(x,p_x)$, $y=0, p_y>0$, Poincar\'{e} surface of section.

It was observed that for a given value of the Jacobi integral $E_{J2}$, a considerable part of the $(x,p_x)$ phase plane is covered by chaotic orbits, when the value of the perturbation strength $\epsilon$ is near the escape perturbation strength $\epsilon_{esc}$. The formula connecting $E_{J2}, \Omega_b$ and $\epsilon_{esc}$ (see Caranicolas \& Karanis, 1998; hereafter CK) is
\begin{flalign}
&\frac{2\epsilon_{esc}\left(19-16\Omega_b^2\right) + 2.5\Omega_b^4 - 11\Omega_b^2 - 2\epsilon_{esc}^2 + 4}
{72\epsilon_{esc}} = E_{J2}.&
\end{flalign}

Fig. 1 shows the $(x,p_x)$ phase plane when $\Omega_b=0, \epsilon=\epsilon_{esc}=1.56113$ and $E_{J2}=0.52$. A considerable part of the phase plane is covered by a chaotic sea. One observes three distinct families of regular orbits (i) orbits producing invariant curves surrounding the central stable invariant point (ii) orbits producing invariant curves that form islands on the $x$ and the $p_x$ axis and (iii) orbits producing a set of small islands embedded on the chaotic sea. In addition to the above orbits there is a family of sticky orbits producing the sticky region around the islands formed by the type (iii) orbits and the family of chaotic orbits forming the large chaotic sea. Fig. 2 is similar to Fig. 1 but when $\Omega_b=0.1, \epsilon=\epsilon_{esc}=1.45597$ and $E_{J2}=0.52$. The only difference with the pattern shown in Fig. 1 is that here the chaotic sea is smaller, while the regular region has increased. This happens because here the perturbation strength is reduced, as $\Omega_b$ has increased and $E_{J2}$ is the same (see also Fig. 13 in CK).
\begin{figure}[!tH]
\resizebox{\hsize}{!}{\rotatebox{0}{\includegraphics*{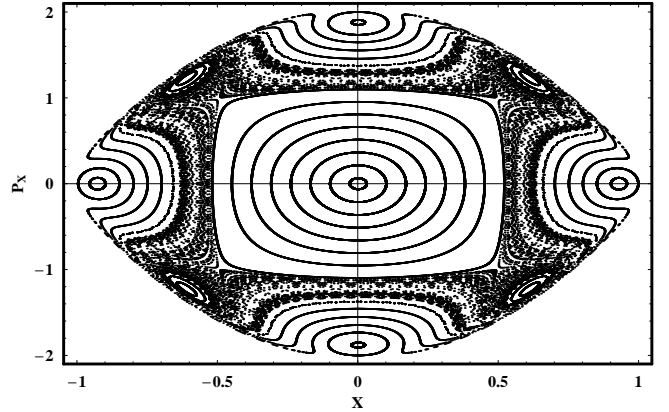}}}
\caption{The $(x,p_x)$ phase plane, when $\Omega_b=0, \epsilon=\epsilon_{esc}=1.56113$ and $E_{J2}=0.52$.}
\end{figure}
\begin{figure}[!tH]
\resizebox{\hsize}{!}{\rotatebox{0}{\includegraphics*{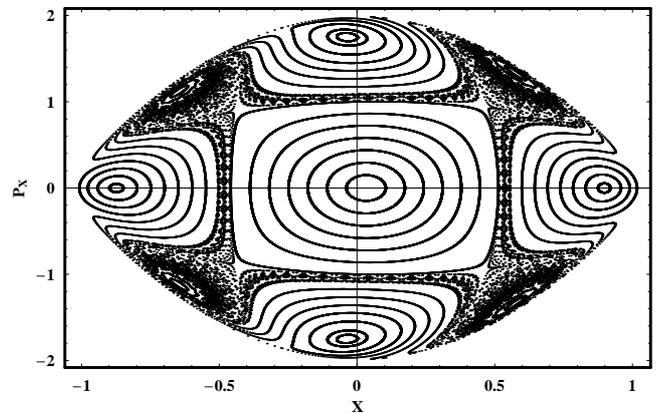}}}
\caption{Similar as Fig.1, but when $\Omega_b=0.1, \epsilon=\epsilon_{esc}=1.45597$ and $E_{J2}=0.52$.}
\end{figure}

Fig. 3a-d shows four representative orbits of the 2D system. Fig. 3a shows an orbit of type (i) when $\Omega_b=0$ and $\epsilon=\epsilon_{esc}=1.56113$. This is a box orbit. The initial conditions are $x_0=0.2, y_0=p_{x0}=0$, while the value of $p_{y0}$ is always found using the Jacobi integral. Fig. 3b shows an orbit of type (ii) when $\Omega_b=0$ and $\epsilon=\epsilon_{esc}=1.56113$. This is quasi periodic orbit starting near the stable figure-eight periodic orbit. The initial conditions are $x_0=0.92, y_0=p_{x0}=0$. Fig. 3c shows an orbits of type (ii) when $\Omega_b=0.1$ and $\epsilon=\epsilon_{esc}=1.45597$. This is quasi periodic orbit producing two small islands symmetric with respect to the center. The initial conditions are $x_0=0.6, y_0=0, p_{x0}=1.28$. A chaotic orbit is given in Fig. 3d. The initial conditions are $x_0=0.58, y_0=p_{x0}=0$, while $\Omega_b=0.1$ and $\epsilon=\epsilon_{esc}=1.45597$. All orbits were calculated for a time period of 100 time units, while $E_{J2}=0.52$.
\begin{figure*}[!tH]
\centering
\resizebox{0.95\hsize}{!}{\rotatebox{0}{\includegraphics*{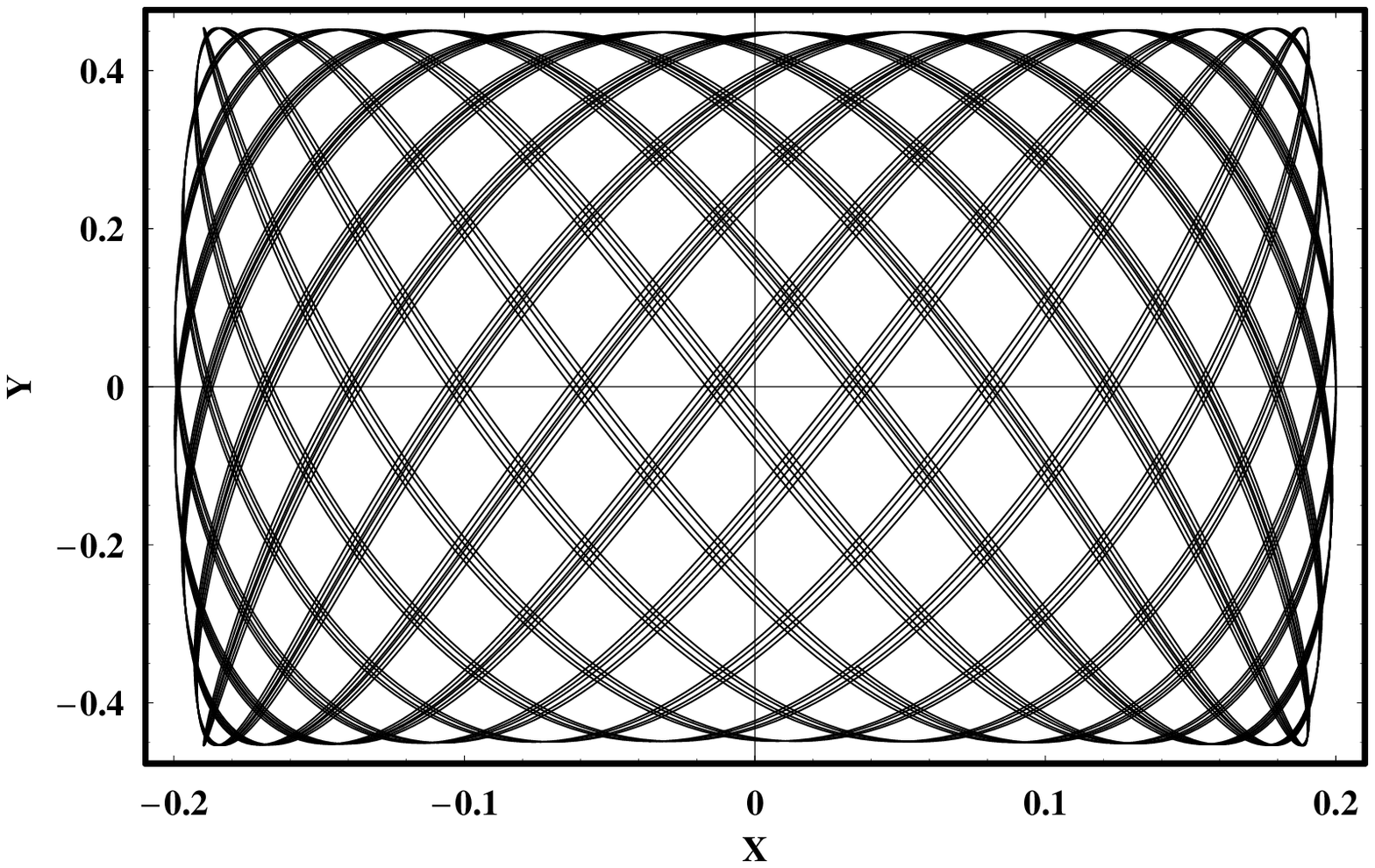}}\hspace{1cm}
                          \rotatebox{0}{\includegraphics*{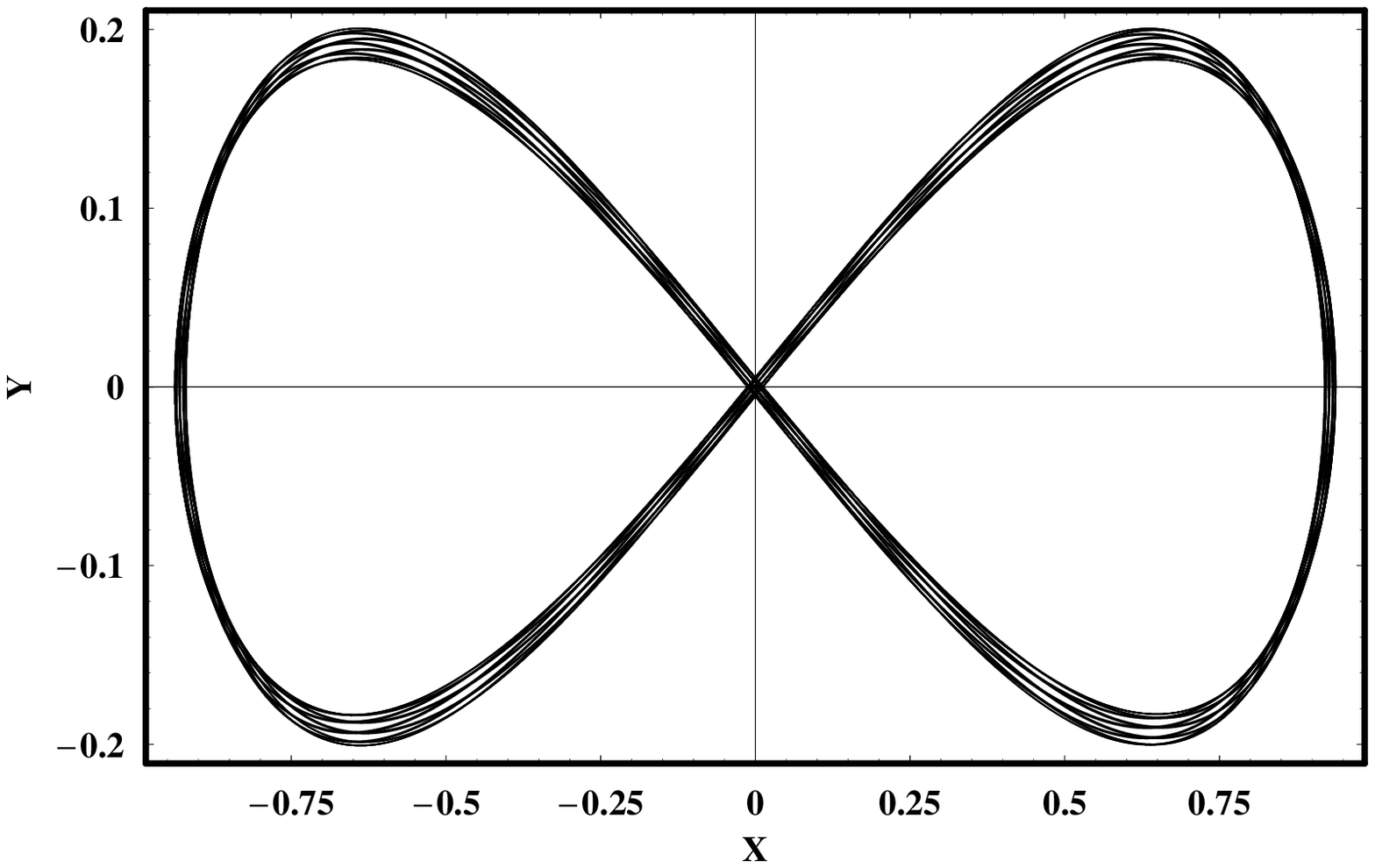}}}
\resizebox{0.95\hsize}{!}{\rotatebox{0}{\includegraphics*{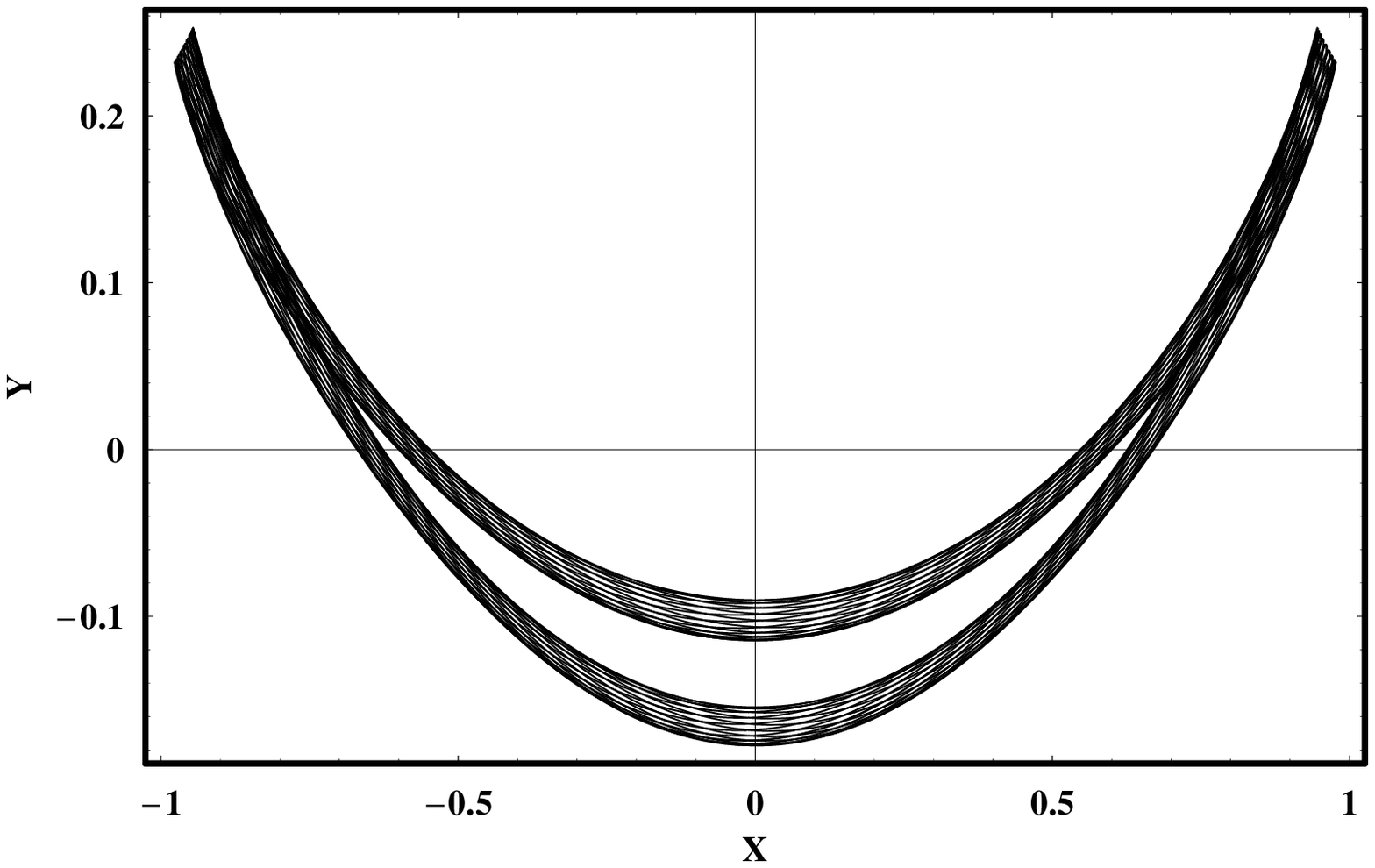}}\hspace{1cm}
                          \rotatebox{0}{\includegraphics*{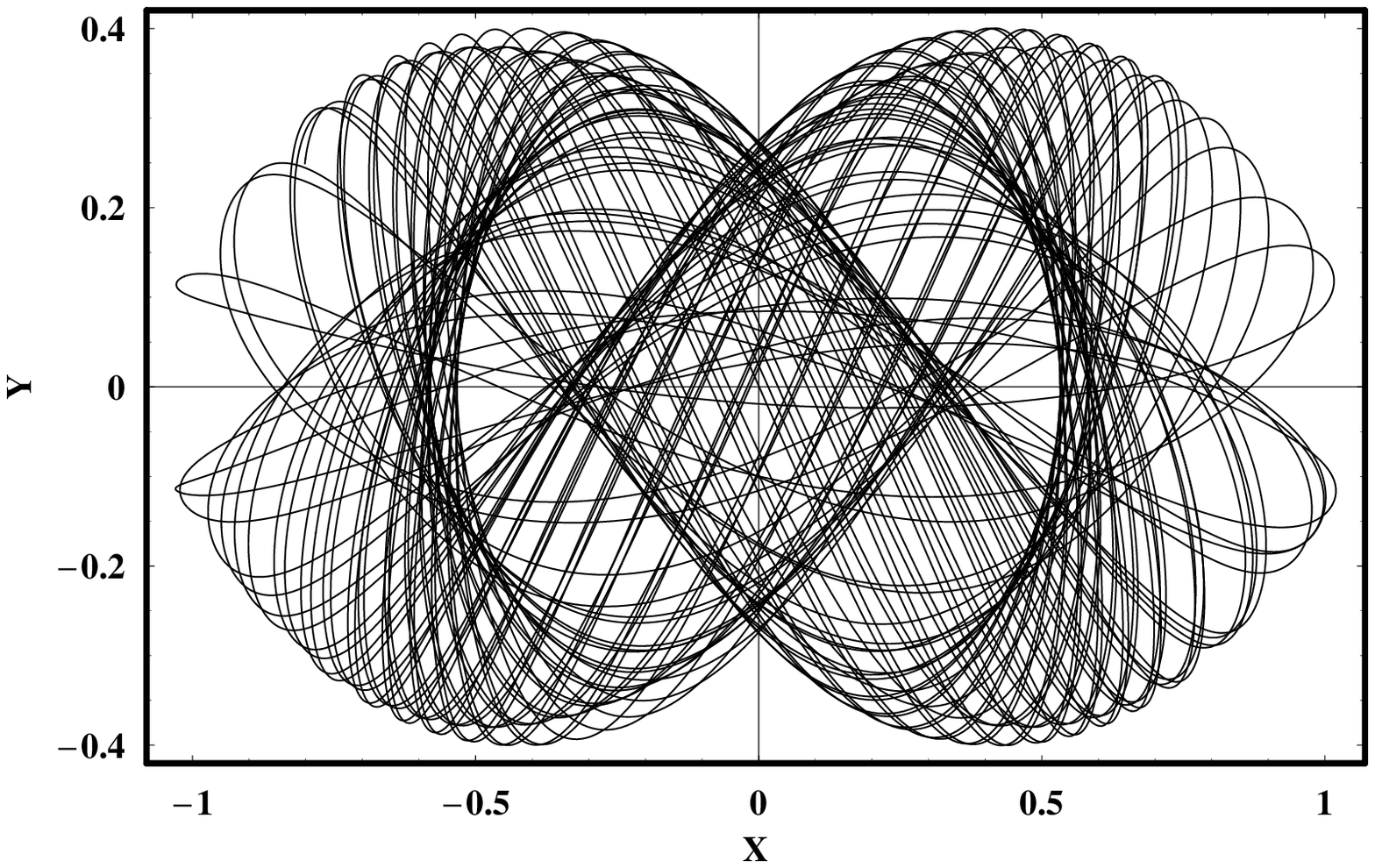}}}
\vskip 0.1cm
\caption{(a)-(d): Orbits in the 2D model. See text for details.}
\end{figure*}
\begin{figure*}[!tH]
\centering
\resizebox{0.95\hsize}{!}{\rotatebox{0}{\includegraphics*{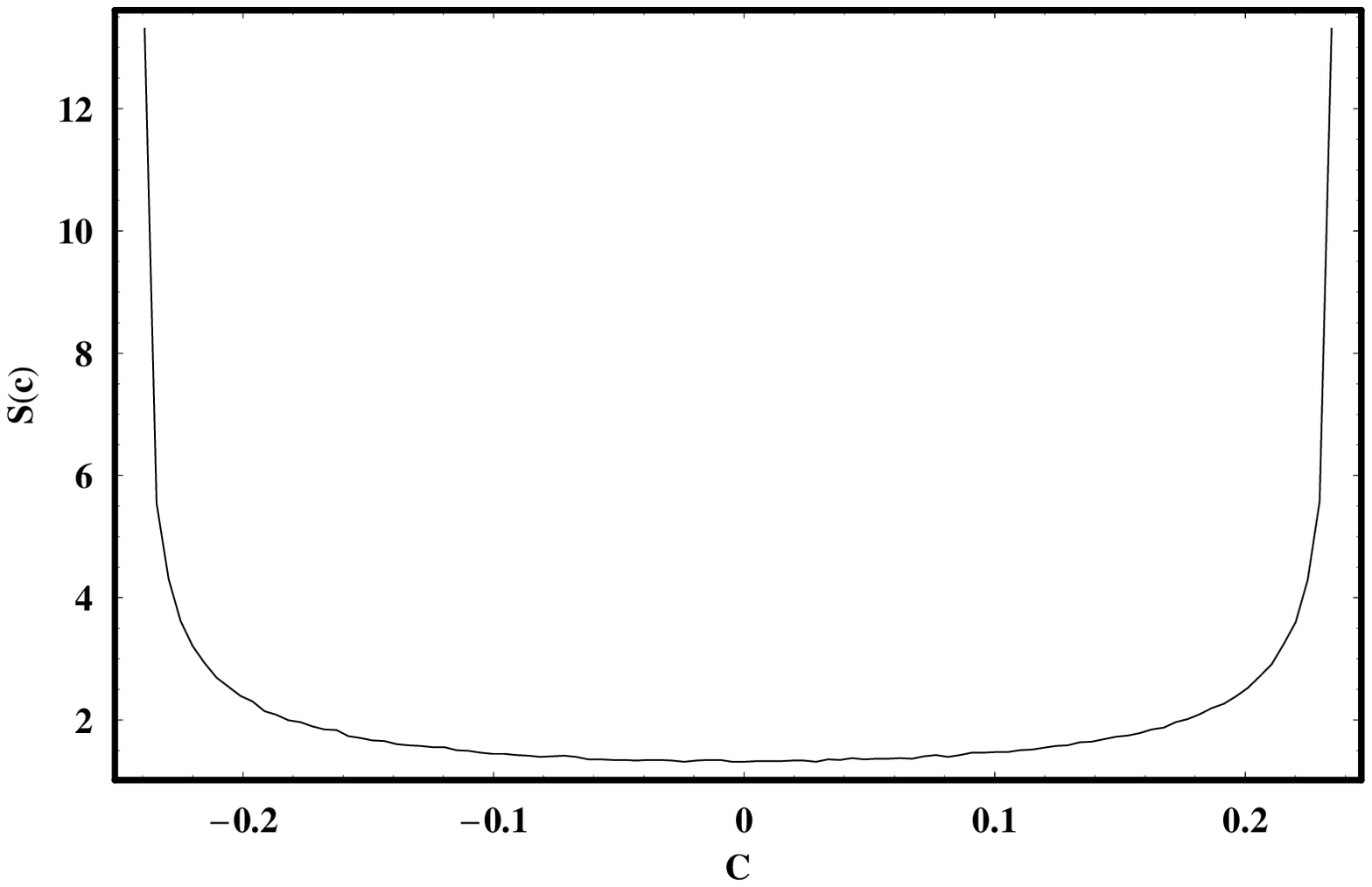}}\hspace{1cm}
                          \rotatebox{0}{\includegraphics*{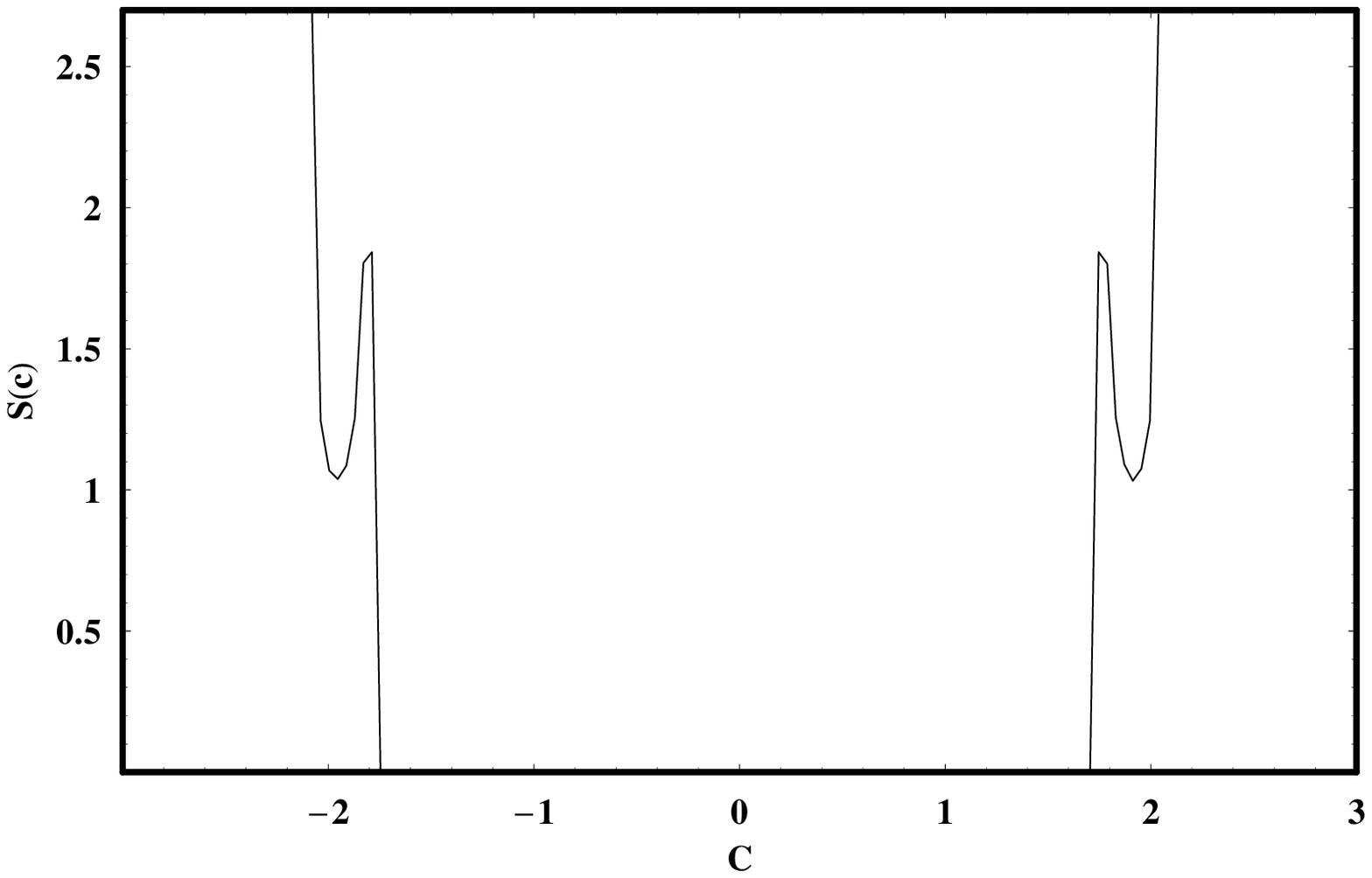}}}
\resizebox{0.95\hsize}{!}{\rotatebox{0}{\includegraphics*{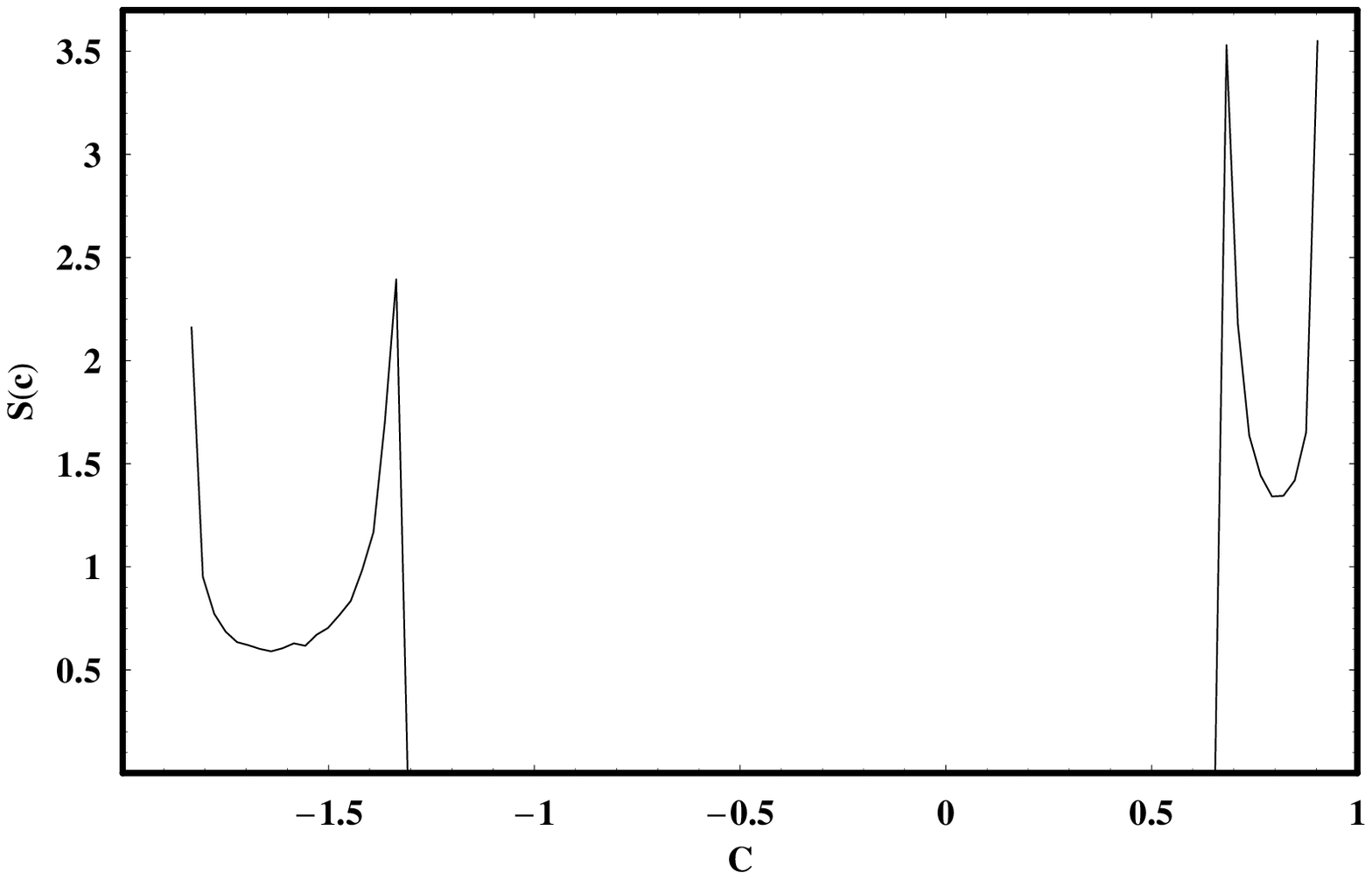}}\hspace{1cm}
                          \rotatebox{0}{\includegraphics*{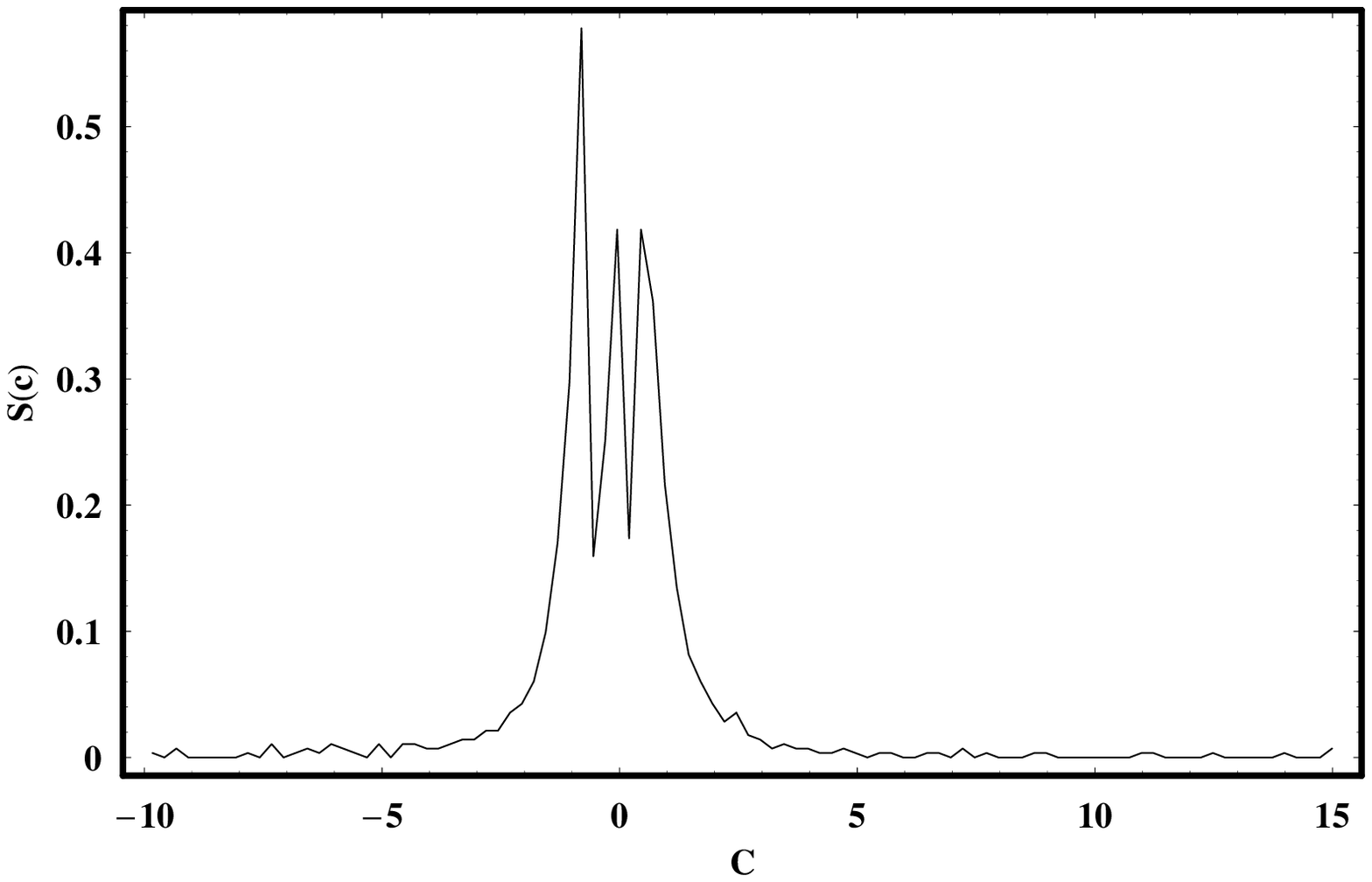}}}
\vskip 0.1cm
\caption{(a)-(d): The $S(c)$ spectra corresponding to the orbits of Fig. 3a-d.}
\end{figure*}
\begin{figure*}[!tH]
\centering
\resizebox{0.95\hsize}{!}{\rotatebox{0}{\includegraphics*{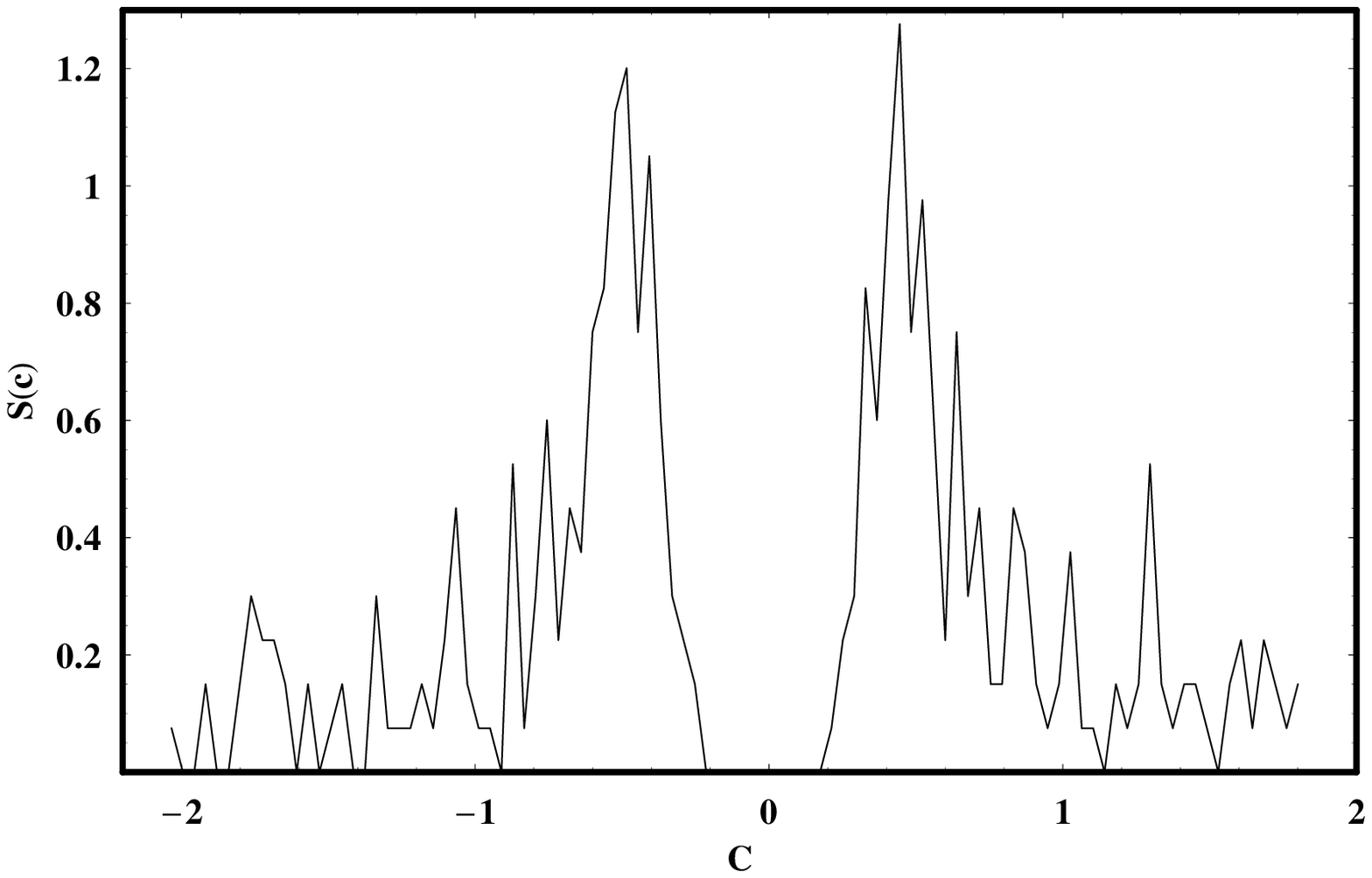}}\hspace{1cm}
                          \rotatebox{0}{\includegraphics*{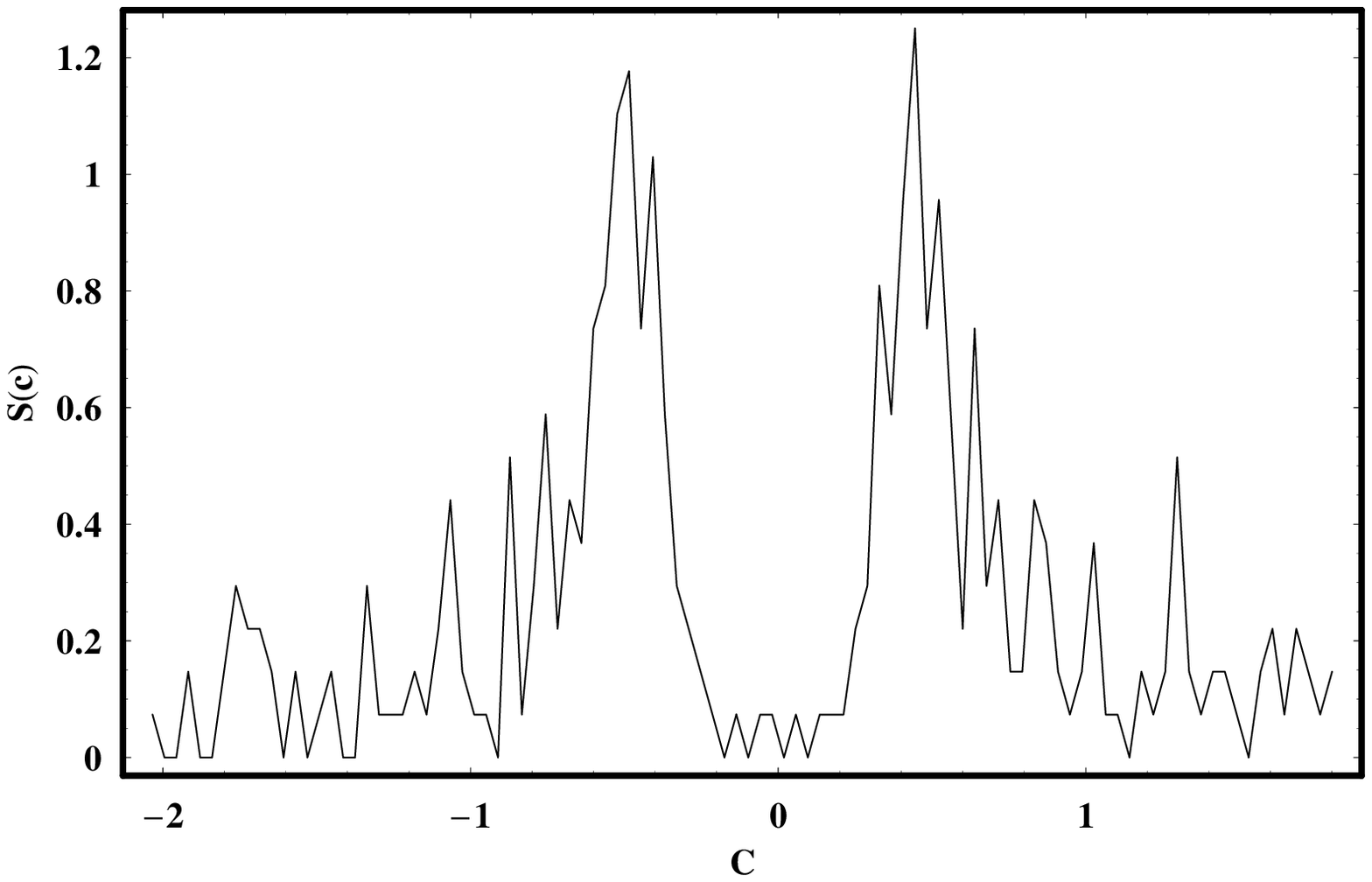}}}
\resizebox{0.95\hsize}{!}{\rotatebox{0}{\includegraphics*{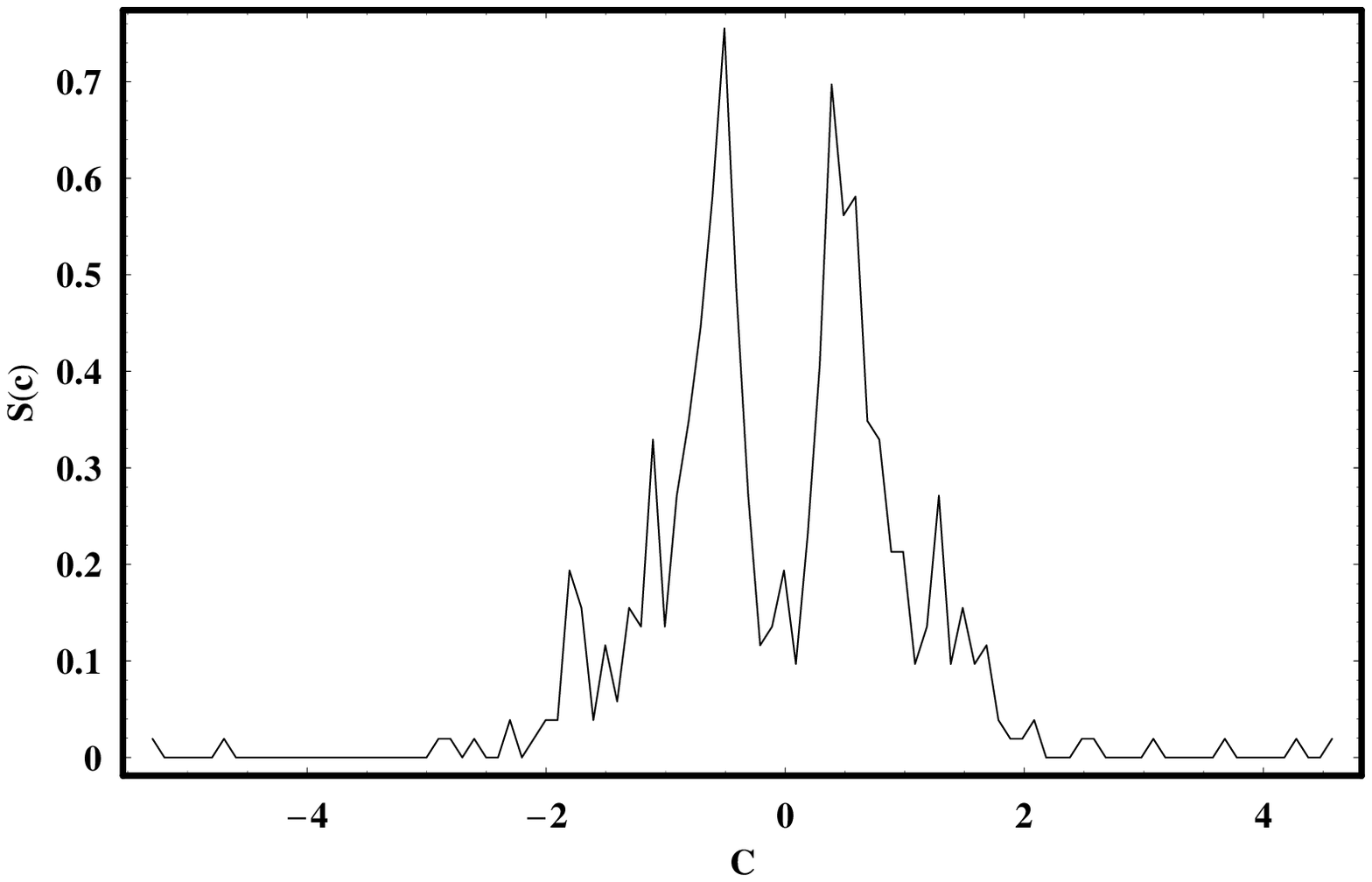}}\hspace{1cm}
                          \rotatebox{0}{\includegraphics*{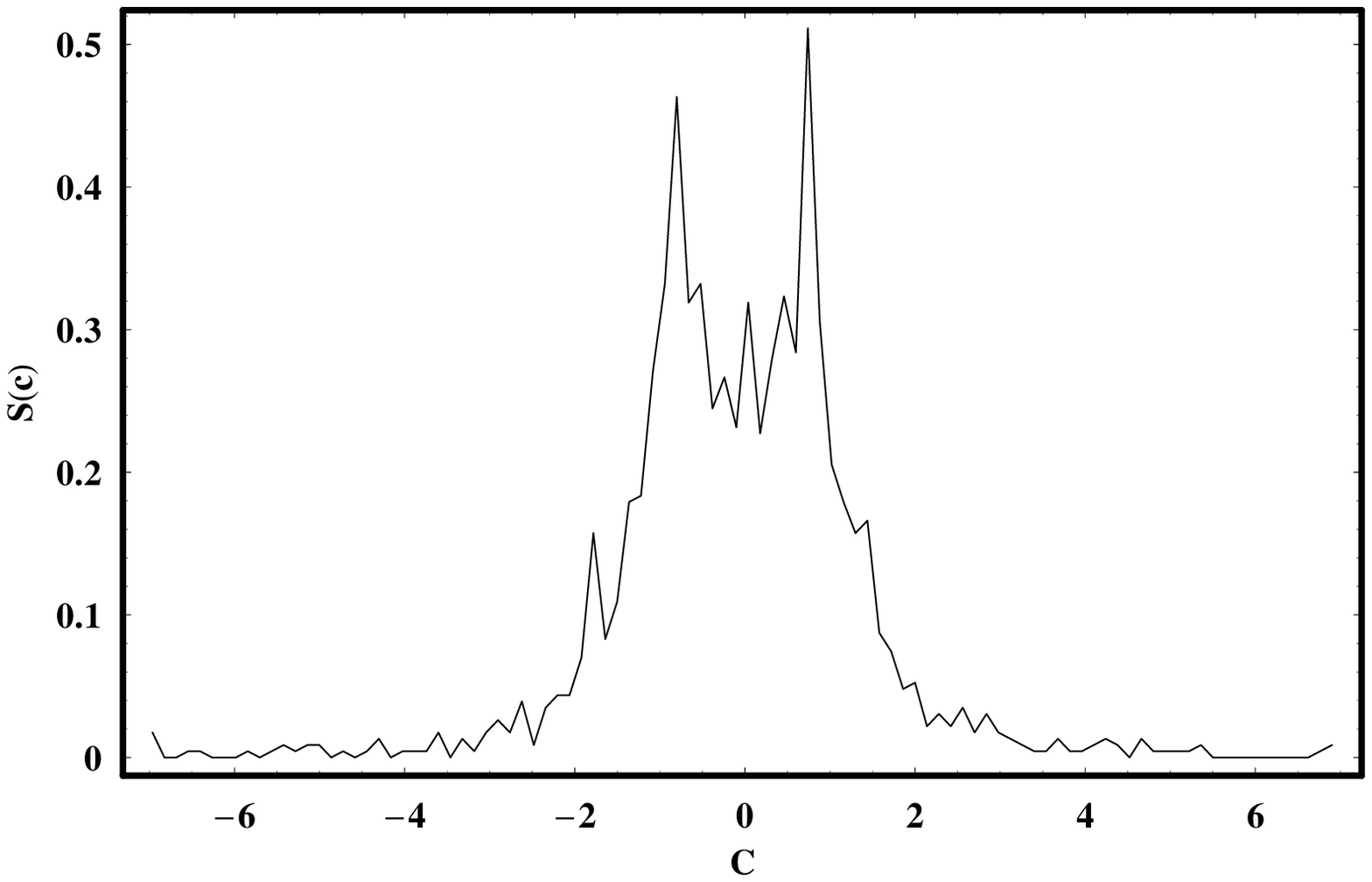}}}
\vskip 0.1cm
\caption{(a)-(d): The evolution of the $S(c)$ spectrum of a sticky orbit. Details are given in the text.}
\end{figure*}
\begin{figure*}[!tH]
\centering
\resizebox{0.95\hsize}{!}{\rotatebox{0}{\includegraphics*{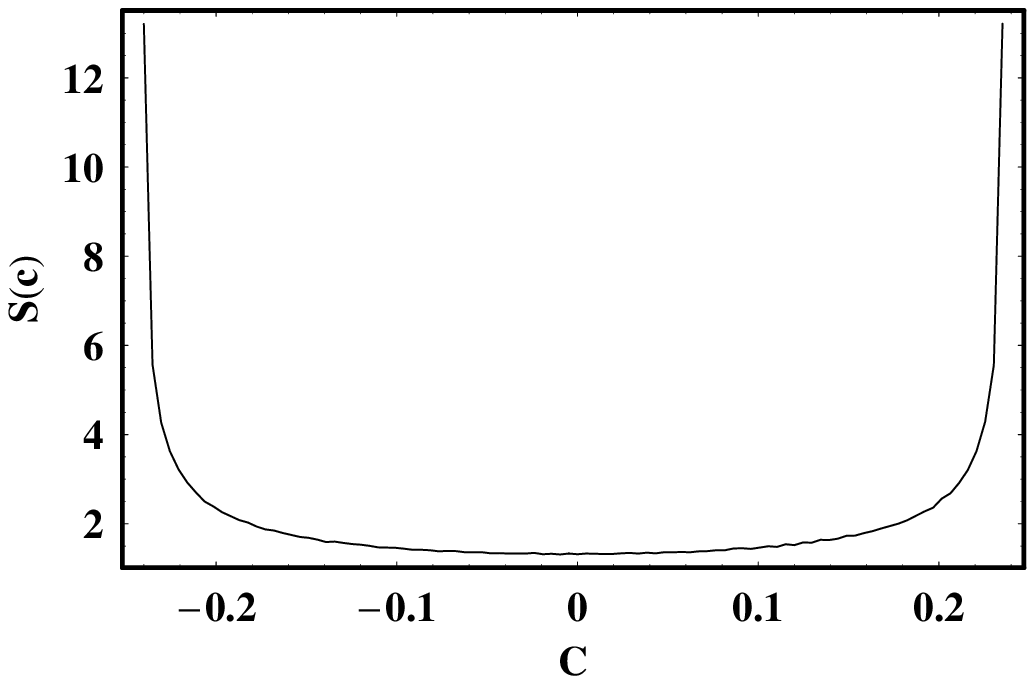}}\hspace{1cm}
                          \rotatebox{0}{\includegraphics*{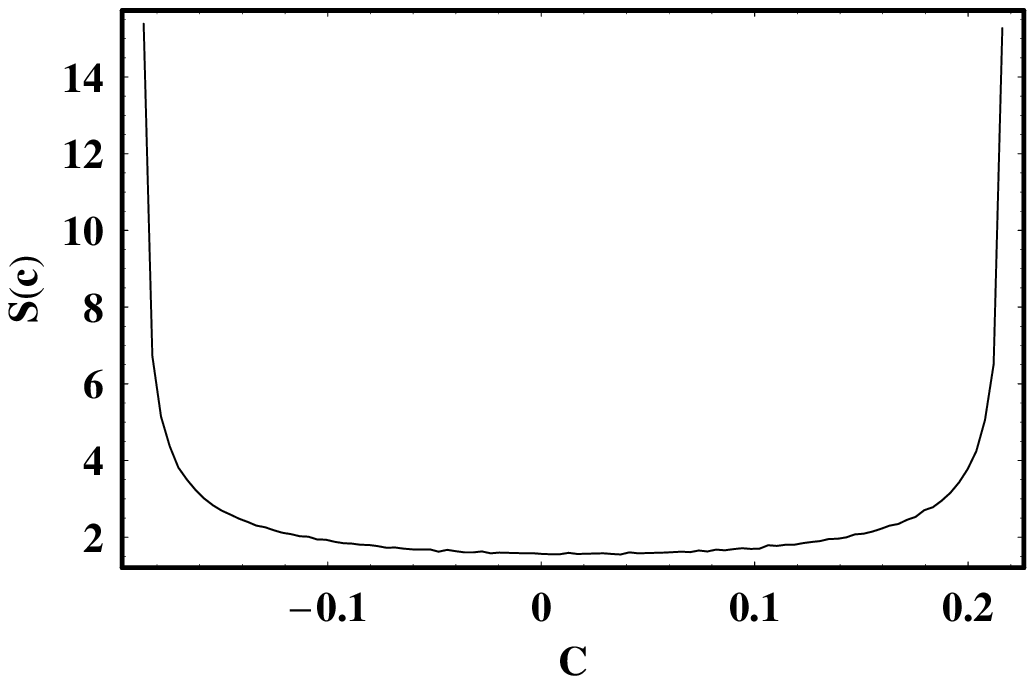}}}
\vskip 0.1cm
\caption{(a)-(b): The $S(c)$ spectrum of two 3D regular orbits. The initial conditions are: $x_0=0.2, y_0=p_{x0}=0, z_0=0.1$. The values of all the other parameters are (a-left) as in Fig. 1 and (b-right) as in Fig. 2.}
\end{figure*}

In the following we shall use the $S(c)$ spectrum in order to distinguish regular motion from chaos. As it was mentioned before, this dynamical parameter was introduced and first used by Caranicolas \& Papadopoulos (2007). The parameter $c_i$ is defined as
\begin{flalign}
&c_i = \frac{x_i - p_{xi}}{p_{yi}},&
\end{flalign}
where $x_i$, $p_{xi}$ and $p_{yi}$ are the successive values of $x$, $p_x$ and $p_y$ on the Poincar\'{e} $(x,p_x)$, $y=0p_y>0$ phase plane. The dynamical spectrum of the parameter $c$ is its distribution function
\begin{flalign}
&S(c) = \frac{\Delta N(c)}{N \Delta c},&
\end{flalign}
where $\Delta N(c)$ are the number of the parameters $c$ in the interval $\left(c, c+\Delta c \right)$, after $N$ iterations. It is clear from Eq. (7) that the $c$ parameter is based on the coordinates and momenta of the test particle (star). The reason for using the $S(c)$ spectrum, apart from distinguishing regular from chaotic motion is: (i) it can identify islandic motion and (ii) it is very useful in order to visualize the sticky motion.

Fig. 4a-d shows the $S(c)$ spectrum for the orbits shown in Fig. 3a-d. Fig. 4a shows the spectrum for the orbit 3a. One can see a well defined U type spectrum indicating regular motion. Fig. 4b shows the spectrum for the orbit given in Fig. 3b. Here we see two well defined U type spectra. This is because the corresponding quasi periodic orbit produces two islands on the $(x,p_x)$ phase plane. Two well defined U type spectra are also produced from the quasi periodic orbit shown in Fig. 3c. The spectrum given in Fig. 4d belongs to the chaotic orbit shown in Fig. 3d. Here things looks very different, as we see a complicated spectrum with large and small peaks. This spectrum is characteristic of chaotic motion.

Here we must note that, our experience from previous work in 2D systems indicates that regular orbits give always U type $S(c)$ spectra, while chaotic orbits give asymmetric spectra displaying large and small peaks. The validity of the results given by the $S(c)$ spectrum was often checked by other indicators, such as the Lyapunov Characteristic Exponent (L.C.E), see Fig. 9 in Caranicolas \& Papadopoulos (2007).

The $S(c)$ spectrum is very useful in order to follow the evolution of the sticky orbits. Fig. 5a-d shows a characteristic example. The corresponding orbit starts in the sticky region near the upper right small island of the surface of section shown in Fig. 1. The initial conditions are $x_0=0.54, y_0=0, p_{x0}=1.23$. During the sticky period, which is about 700 time units one observes, in Fig. 5a, two separate very complicated spectra with a large number of asymmetric peaks. In Fig. 5b, the time is 710 time units and the two spectra are very similar to those given in Fig. 5a. Note that here the two spectra are connected. This indicates that the test particle (star) has left the sticky region in order to continue its wandering in the chaotic region. Fig. 5c shows the spectrum for 1000 time units. We see that the spectrum tends to take the characteristics of a chaotic spectrum. Finally in Fig. 5d we see the spectrum for 3000 time units. Here the spectrum has the characteristic of a spectrum produced by a chaotic orbit.

From the above analysis it is clear that the $S(c)$ spectrum is a useful tool in order to investigate the character of motion in the 2D system. In the next Section, we shall show that this dynamical parameter is also useful in order to distinguish between order and chaos in the 3D dynamical model as well.

\section{Orbits and spectra in the 3D model}

In this Section, we shall try to investigate the character of motion in the 3D potential (3). In order to obtain this we shall use the results of the 2D system. We use initial conditions $(x_0,p_{x0},z_0)$,$y_0=p_{z0}=0$, where $(x_0,p_{x0})$ is a point on the phase plane of the 2D system. This point lies inside the limiting curve, that is the curve containing all the invariant curves of the 2D system. This is given by the equation
\begin{flalign}
&\frac{1}{2}p_x^2 + V_{eff}\left(x\right) = E_{J2}.&
\end{flalign}

We take $E_J=E_{J2}$ and the value of $p_{y0}$ for all orbits is obtained from the Jacobi integral (4).

In order to obtain the 3D $S(c)$ spectrum of an orbit, we take the sections of the 3D orbit with the plane $y=0$, when $p_y>0$. The set of the four dimensional points $(x,p_x,z,p_z)$ is projected on the $(x,p_x)$ plane and we take the $S(c)$ spectrum using relations (7) and (8). The main difference here, is that the above procedure involves also the coupling of the third component $z$.

All numerical experiments indicate that orbits with initial condition such as $(x_0,p_{x0})$ is a point in the chaotic sea of Figs. 1 and 2 and for all permissible values of $z_0$ give chaotic orbits. On the other hand, it would be also interesting to know what happens with the orbits with initial conditions such as $(x_0,p_{x0})$ is a point in the regular area of Figs. 1 and 2. In order to do this we take the $S(c)$ spectrum of the corresponding 3D orbit.
\begin{figure*}[!tH]
\centering
\resizebox{0.95\hsize}{!}{\rotatebox{0}{\includegraphics*{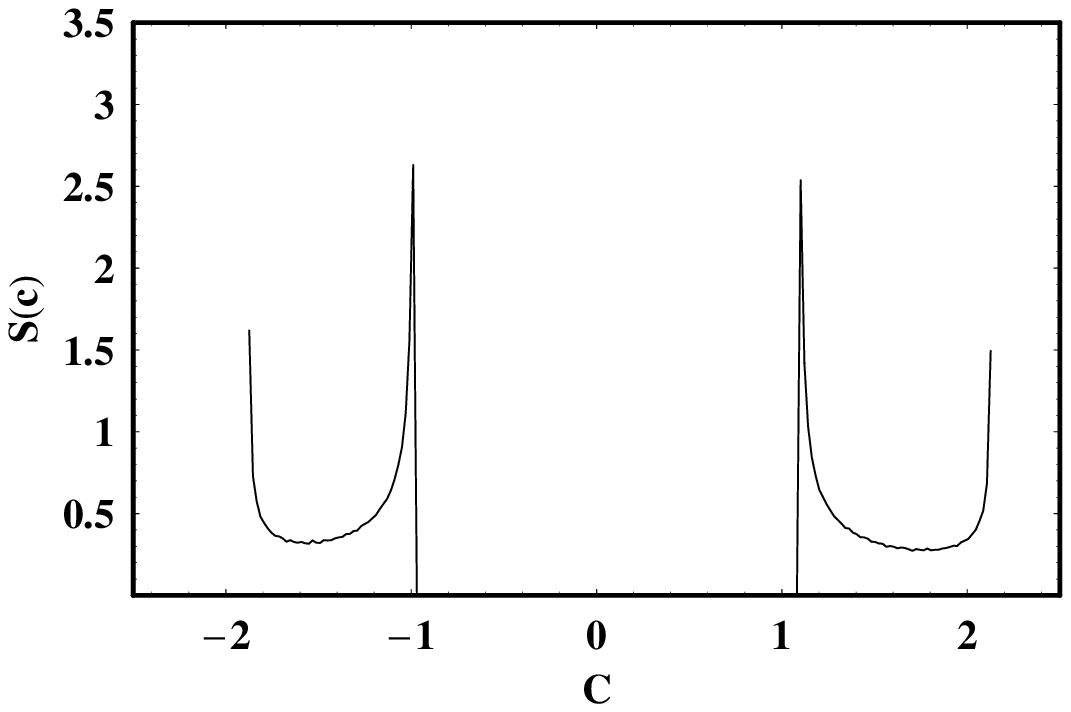}}\hspace{1cm}
                          \rotatebox{0}{\includegraphics*{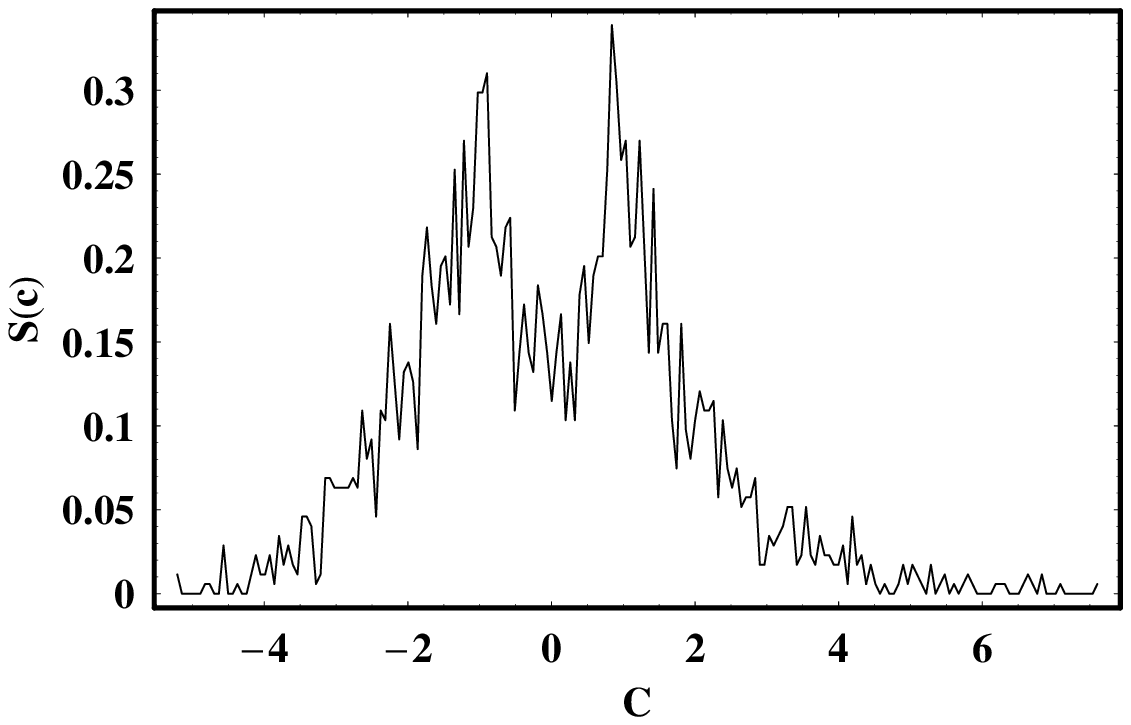}}}
\vskip 0.1cm
\caption{(a)-(b): The $S(c)$ spectrum of two 3D regular orbits. The initial conditions are: (a-left) $x_0=0.85, y_0=p_{x0}=0, z_0=0.02, \Omega_b=0.1$ and (b-right) $x_0=0.85, y_0=p_{x0}=0, z_0=0.3, \Omega_b=0.1$.}
\end{figure*}
\begin{figure*}[!tH]
\centering
\resizebox{0.95\hsize}{!}{\rotatebox{0}{\includegraphics*{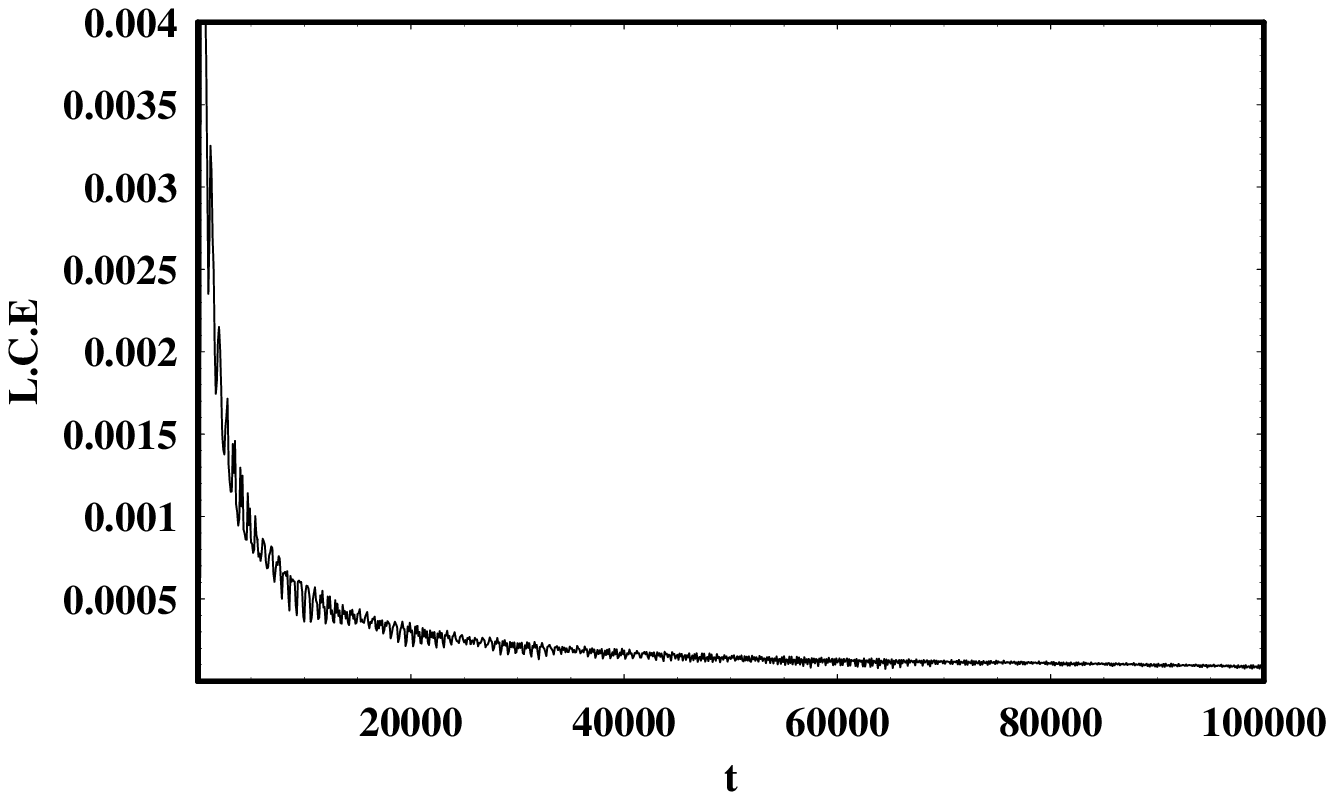}}\hspace{1cm}
                          \rotatebox{0}{\includegraphics*{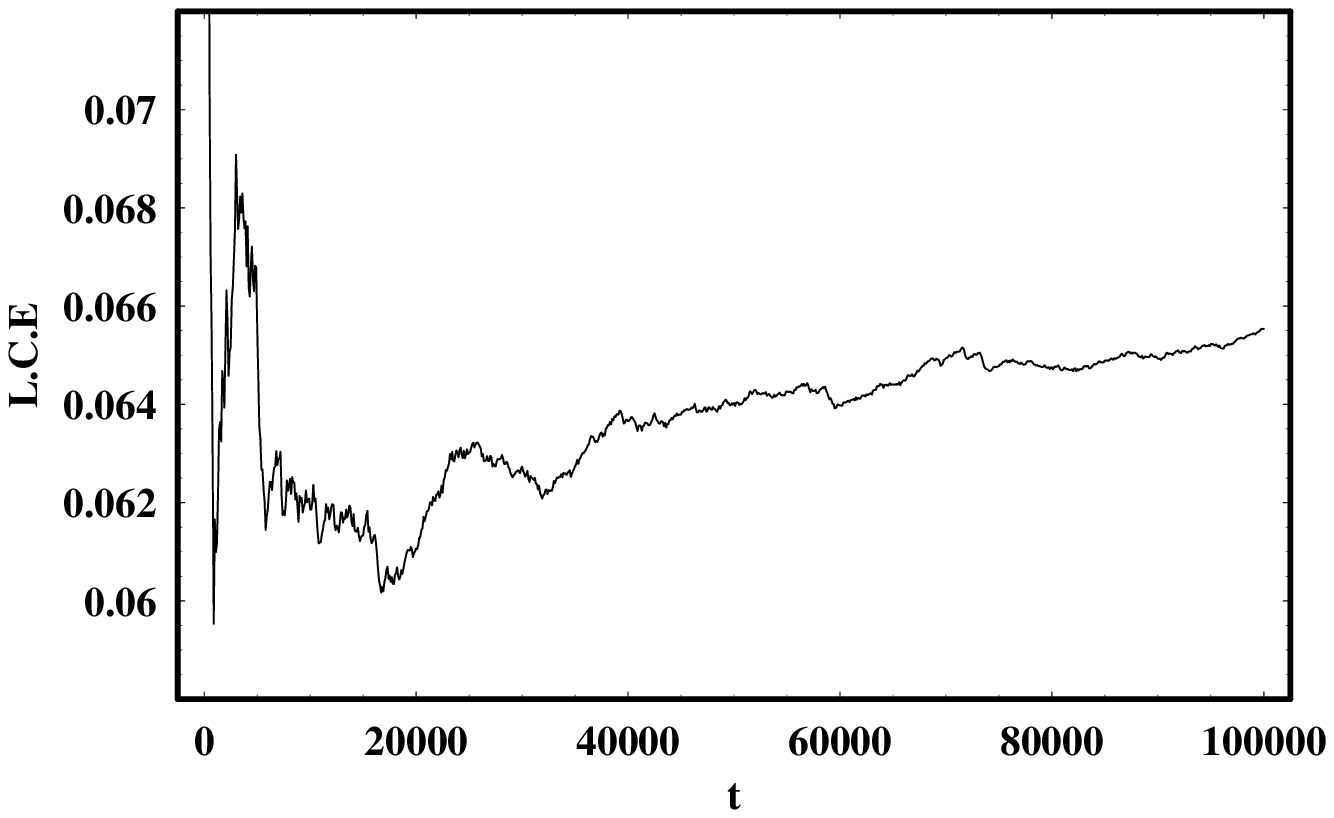}}}
\vskip 0.1cm
\caption{(a)-(b): (a-left): The L.C.E for the regular orbit 7a and (b-right): The L.C.E for the chaotic orbit 7b.}
\end{figure*}
\begin{figure*}[!tH]
\centering
\resizebox{0.75\hsize}{!}{\rotatebox{0}{\includegraphics*{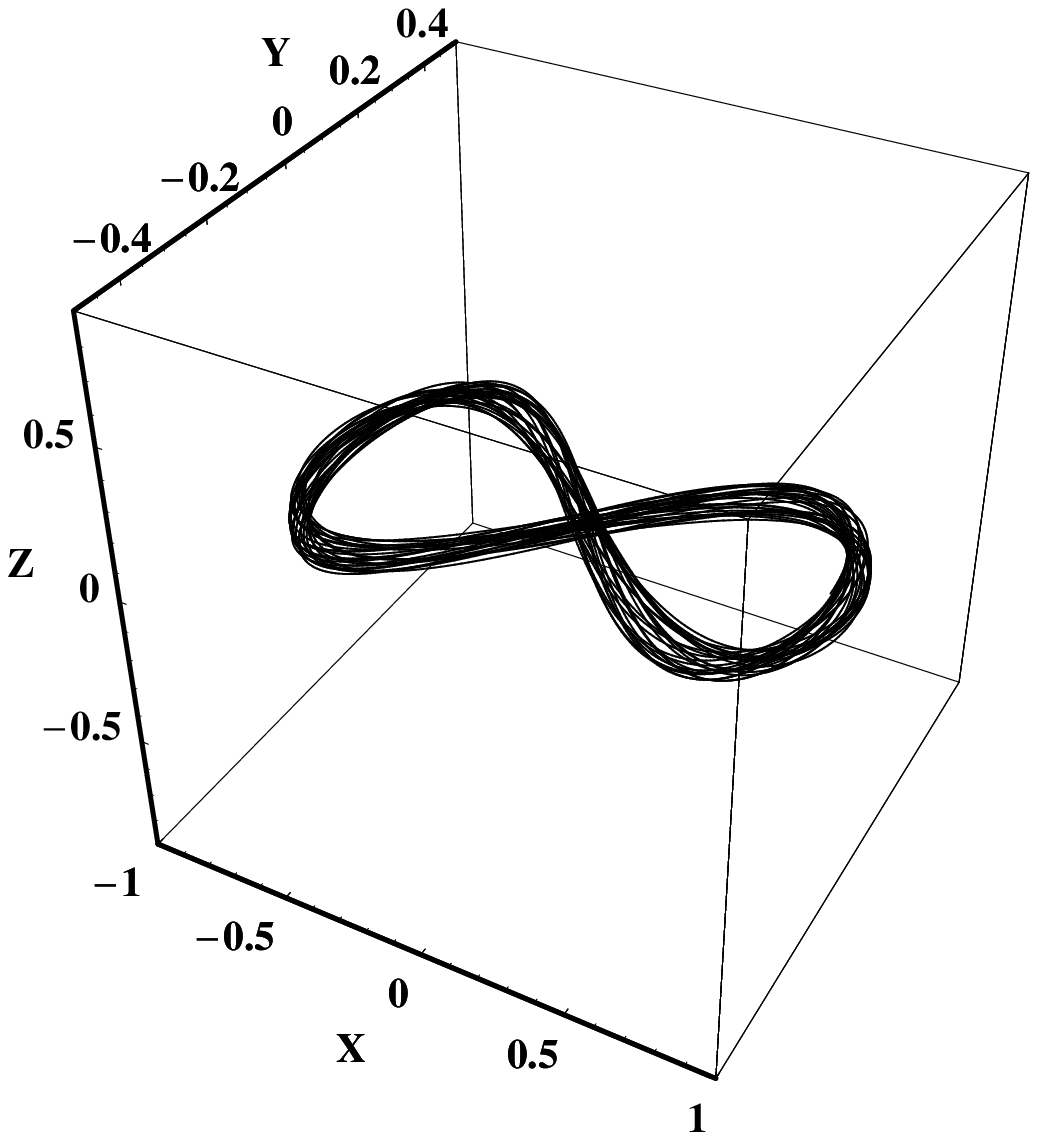}}\hspace{1cm}
                          \rotatebox{0}{\includegraphics*{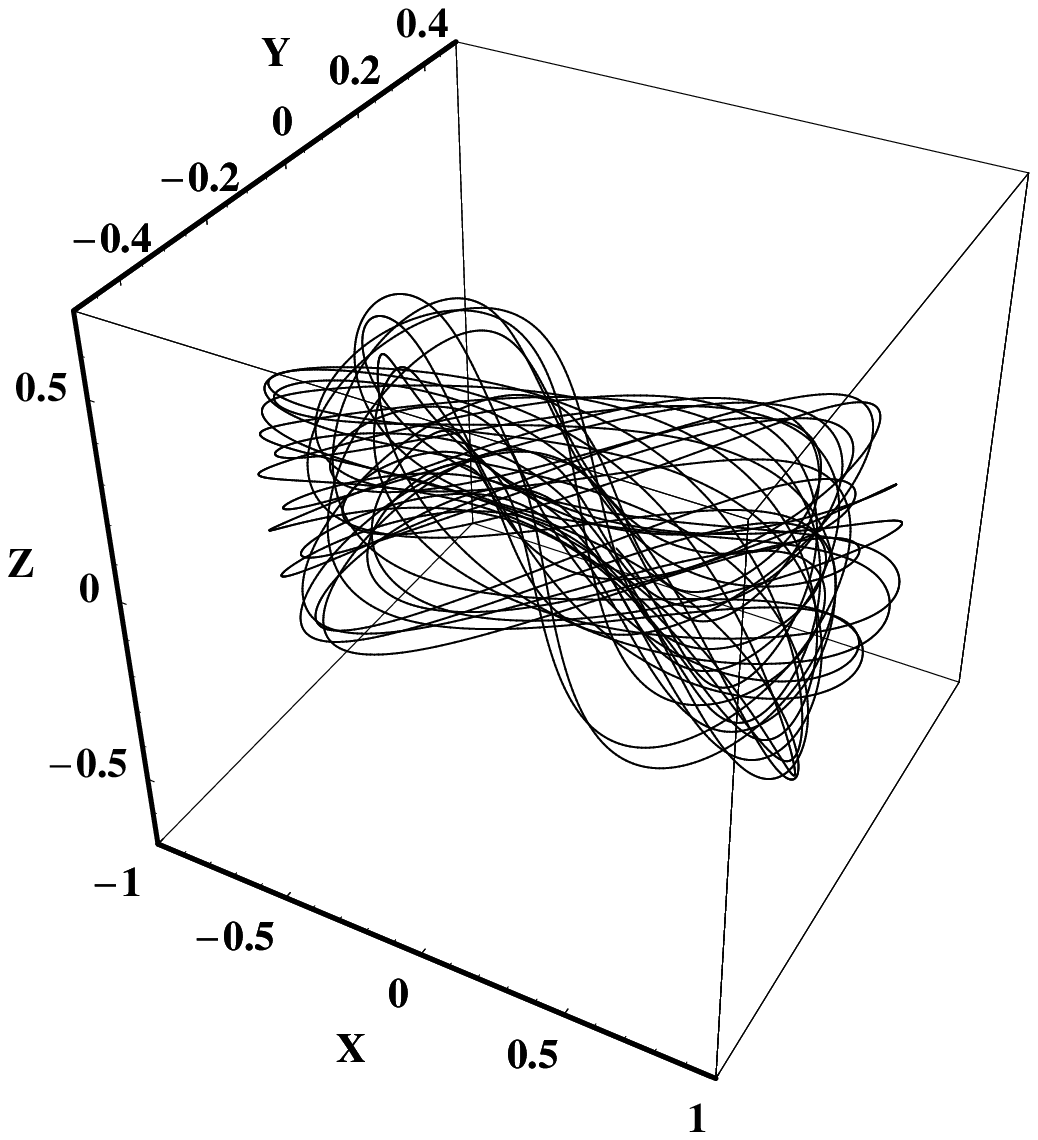}}}
\vskip 0.1cm
\caption{(a)-(b): The orbits giving the spectra in Fig. 7a and b respectively.}
\end{figure*}

Fig. 6a-b shows the $S(c)$ spectrum of two 3D orbits. In Fig. 6a the values of all parameters are as in Fig. 1, while the initial conditions of the orbit are $x_0=0.2, y_0=p_{x0}=0, z_0=0.1$. The corresponding spectrum is a well defined U type spectrum indicating regular motion. The orbit shown in Fig. 6b has the same initial conditions as that of Fig. 6b but the values of the parameters are as in Fig. 2. Again the spectrum indicates regular motion. A large number of spectra were calculated for orbits in the regular region around the central invariant point of Figs. 1 and 2 and for all permissible values of $z_0$ (up to about 1.5) show that the motion is regular.

The second and more interesting step is to see what happens for orbits starting in the regular region and particularly near the islands produced by the figure-eight orbits. This is important because these orbits are the orbits that have generated this potential and are those supporting the barred structure. On this basis, it is interesting to see if they retain their regular character in the 3D potential. A characteristic example is given in Fig. 7a and b. Fig. 7a shows the spectrum of an orbit with initial conditions $x_0=0.85, y_0=p_{x0}=0, z_0=0.02$. The values of all the other parameters are as in Fig. 2. As this spectrum looks similar to that produced by the two-dimensional figure-eight orbit in Fig. 4c, one could say that this spectrum is produced by a three-dimensional figure-eight orbit. We shall come to this point later in this Section. Fig. 7b shows the spectrum of an orbit with all the other parameters as those of Fig. 7a but when $z_0=0.3$. Here, instead of two U shaped spectra, one can see an asymmetric spectrum with a lot of small and large peaks. This spectrum suggests that the corresponding orbit is chaotic.

In order to have a second indicator about the regular or chaotic character of the 3D orbits giving the spectra shown in Fig. 7a and b, we computed the L.C.E for each orbit, for a time period of $10^{5}$ time units. The results are shown in Fig. 8a and b respectively. We can see that the results of the $S(c)$ spectrum coincide with the outcomes given by the L.C.E.

Fig. 9a and b shows the 3D orbits giving the spectra shown in Fig. 7a and b respectively. In Fig. 9a we see a 3D figure-eight quasi periodic orbit. Note that the orbit stays close to the galactic plane. On the contrary, the orbit shown in Fig. 9b has lost its figure-eight structure. Strictly speaking this orbit is a chaotic orbit and it goes to considerably larger values of $z$. Note that both the above orbits support the barred structure. Therefore, we can conclude that near the galactic plane the 3D barred structure is supported by regular orbits, while in higher values of $z$ the barred structure is supported by chaotic orbits. Furthermore, it is evident that the 3D spectra provide excellent evidence about the regular or chaotic character of the 3D orbits.

\section{Discussion and conclusions}

Galactic potentials made up of harmonic oscillators have been extensively used for about five decades in order to model motion in galaxies (see H\'{e}non \& Heiles, 1964; Deprit, 1991; Lara, 1996; Deprit \& Elipe, 1999; Elipe, 2000; Arribas et al., 2006; Karanis \& Vozikis, 2008). In all the above models the choice of the parameters was more or less arbitrary. On the contrary, our model is a perturbed 3D harmonic oscillator and it is an extension to the 3D space of the potential obtained using a family of figure-eight orbits and the theory of the inverse problem of dynamics (IPD). It is worth mentioning that the theory of the inverse problem of dynamics has made significant progress during the last years with interesting results (see e.g. Anisiu \& Bozis, 2004; Anisiu, 2005; Bozis \& Blaga, 2006; Bozis \& Kotoulas, 2006).

In this article, we have studied the regular or chaotic character of motion in a 3D dynamical model describing the motion in the central parts of a barred galaxy. In order to do this we started from the $(x,p_x)$ phase plane of the 2D system. As chaotic motion is observed for large values of $\epsilon$, we have used a fixed values of $E_{J2}$ and $\Omega_b$ and $\epsilon=\epsilon_{esc}$.

In the 2D system, as expected, the regular or chaotic character of orbits, the evolution of the sticky orbits and the islandic motion is very well explored using the $S(c)$ spectrum. Moreover, the $S(c)$ spectrum has given interesting information for the 3D system as well. The numerical calculations have shown that all orbits starting inside the chaotic sea of the 2D system give spectra indicating chaotic motion. On the other hand, the spectra of orbits starting in the regular region of Figs. 1 and 2 indicate regular or chaotic motion. The numerical calculations suggest, that 3D orbits starting near the central invariant point give always U type spectra for all permissible values of $z_0$. Thus, one can say that the motion near the center of the system is regular. The situation is quite different for the figure-eight orbits. These orbits retain their figure-eight shape for small values of $z_0$, $(z_0 \lesssim 0.05)$, while for large values of $z_0$, $(z_0 > 0.05)$ they loose their shape and become chaotic.

We must emphasize that the $S(c)$ spectrum of the 3D orbits is computed taken into account all coordinates and momenta $(x,p_x,z,p_z)$ and $p_y>0$ with $y=0$, as described in Section 3. Of course, Eq. (7) does not contain $z$ and $p_z$ but the coupling of the $z$ components, carrying all information about the 3D motion, is hidden in the values of $x$, $p_x$ and $p_y$ of the 3D orbit. Furthermore, comparison with the L.C.E shows that the results given by the $S(c)$ spectrum are completely reliable.

The above results lead to the conclusion that the $S(c)$ spectrum is very useful in order to distinguish the regular or chaotic motion in 3D potentials. Furthermore, it can indicates 3D quasi periodic orbits (see Fig. 8a). The $S(c)$ spectrum is faster than the $S(\alpha)$ spectrum used in Caranicolas \& Vozikis (1999), because needs only one orbit and a small number of iterations $10^3-10^4$. Moreover, the $S(\alpha)$ spectrum has no ability to detect islandic and sticky motion, as the $S(c)$ spectrum does.

\section*{Acknowledgement}

The authors would like to thank the anonymous referee for his useful suggestions and comments.

\end{document}